\documentclass[%
 reprint,
 amsmath,amssymb,
 aps,
]{revtex4-2}

\usepackage{graphicx}% Include figure files
\usepackage{dcolumn}% Align table columns on decimal point
\usepackage{bm}% bold math
\usepackage{diagbox}
\begin{document}

\preprint{APS/123-QED}

\title{Polarization-encoded quantum key distribution with a room-temperature telecom single-photon emitter}% Force line breaks with \\

\author{Zhang Xingjian\textsuperscript{1}}
\email{xingjianzhang@u.nus.edu}
\author{Zhang Haoran\textsuperscript{2}}
\email{haoran.zhang@ntu.edu.sg}
\author{Rui Ming Chua\textsuperscript{1,3}}
\author{John Eng\textsuperscript{2}}
\author{Max Meunier\textsuperscript{2}}
\author{James A Grieve\textsuperscript{3}}
\author{Gao Weibo\textsuperscript{1,2,4}}
\email{wbgao@ntu.edu.sg}
\author{Alexander Ling\textsuperscript{1,5}}
\email{phyalej@nus.edu.sg}

\affiliation{\textsuperscript{1}Centre for Quantum Technologies, 3 Science Drive 2, National University of Singapore, Singapore 117543, Singapore}
\affiliation{\textsuperscript{2}Division of Physics and Applied Physics, School of Physical and Mathematical Sciences, Nanyang Technological University, Singapore 637371, Singapore}
\affiliation{\textsuperscript{3}Quantum Research Center, Technology Innovation Institute, Masdar City, Abu Dhabi, UAE}
\affiliation{\textsuperscript{4}School of Electrical and Electronic Engineering, Nanyang Technological University, Singapore 637371, Singapore}
\affiliation{\textsuperscript{5}Department of Physics, National University of Singapore, Singapore 119077, Singapore}

\date{\today}% It is always \today, today,

\begin{abstract}
Single photon sources (SPSs) are directly applicable in quantum key distribution (QKD) because they allow the implementation of the canonical BB84 protocol. To date, QKD implementations using SPS are not widespread because of the need for cryogenic operation, or frequency conversion to a wavelength efficiently transmitted over telecommunication fibers. We report an observation of polarization-encoded QKD using a room-temperature telecom SPS based on a GaN defect. A field test over 3.5~km of deployed fiber with 4.0 dB loss yielded a secure key rate of 585.9~bps. Further testing in a 32.5 km fiber spool (attenuation of 11.2 dB), which exhibited substantially lower polarization mode dispersion, yielded a secure key rate of 50.4 bps. Both results exhibited a quantum bit error rate (QBER) of approximately 5\%. These results illustrate the potential of the GaN defects for supporting polarization-encoded quantum communication.
\end{abstract}

\maketitle

%\tableofcontents

\section{\label{sec:level1}Introduction}

The original quantum key distribution (QKD) protocol by Bennett and Brassard (BB84) proposed that two parties sharing single photons could generate a secure encryption key~\cite{Bennett2014}. Due to their ease of implementation, many QKD implementations rely on weak coherent pulses~\cite{inoue2003differential,deng2004bidirectional,stucki2005fast} using decoy state techniques~\cite{Lo05prl,wang2005beating,hwang2003quantum}. Although impressively high key rates and long distances~\cite{liu20231002} have been achieved with these sources over fiber and free space, performing QKD with single photon sources (SPS) remains appealing due to reduced engineering overhead, and a reduction in potential security loopholes~\cite{lo2005phase,Tang2013}.
%%Not sure I buy the argument that single photons are needed for device independent QKD, and the potential for fully device-independent QKD\cite{vazirani2019fully}.

%Setup Fig here simply to make it correct place
\begin{figure*}[htbp]
\includegraphics[scale=0.4]{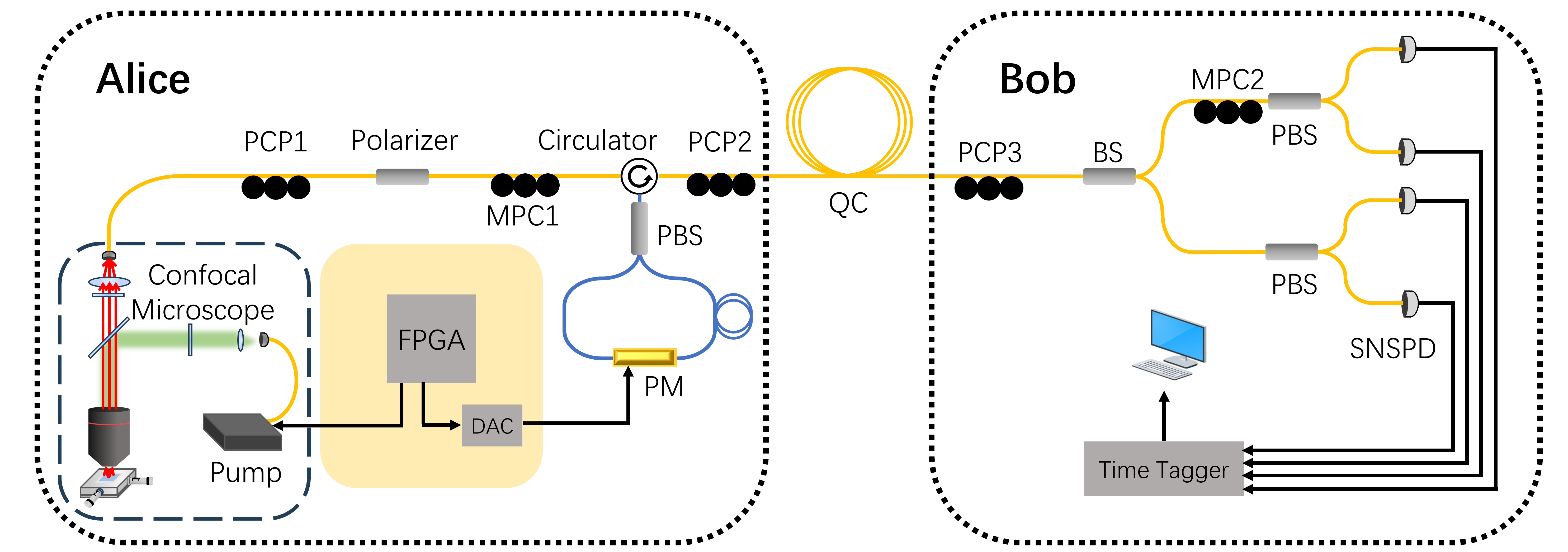}
\centering
\caption{Experimental setup for SPS BB84. PCP: Polarization Controller Paddles. MPC: Motorized Polarization Controller. PM: Phase Modulator. QC: Quantum Channel. BS: Beam Splitter. PBS: Polarization Beam Splitter. SNSPD: Super-conducting Nanowire Single Photon Detector. DAC: Digital-Analog Converter. Yellow lines represent single-mode fiber, while blue ones are Polarization-Maintaining fiber. }
\label{setup}
\end{figure*}

Many materials have been investigated as solid-state SPSs. Examples include semiconductor quantum dots~\cite{senellart2017high,claudon2010highly,maier2014bright,birowosuto2012fast} and nitrogen-vacancy color centers~\cite{rodiek2017experimental,schroder2011fiber}. Reliable telecom-band SPSs are particularly desirable for QKD applications due to their compatibility with existing deployed fiber links. However, most SPSs emit at wavelengths shorter than the telecom band~\cite{Seni21, Murtaza:23, heindel2012quantum}, where dedicated fibers~\cite{zhang2024experimental} or additional frequency conversion steps~\cite{zahidy2024quantum,morrison2023single} are required. Moreover, many telecom-band SPSs require cryogenic cooling systems~\cite{gyger2022metropolitan,takemoto2015quantum,ward2005demand}, limiting their commercial feasibility for QKD. A GaN-based SPS operating at room temperature within the telecom band\cite{Yuzhou18} is therefore an excellent quantum source for metropolitan QKD~\cite{haoran2024}. Alternatively, SPSs outside the telecom band, such as hBN, which also operate at room temperature, could enable QKD over free-space links~\cite{Zeng22}, although they are not compatible with fiber-based communication.

We report a polarization-encoded QKD implementation using a room temperature, telecom-compatible SPS based on a GaN-defect. The emitter produces single photons centered at 1309.5~nm (which is in the telecom O-band) and was used to demonstrate the BB84 protocol. A field trial was first performed over a 3.5~km deployed fiber loop with a loss of 4.0 dB. By selecting the appropriate polarization states for transmission, it was possible to minimize the effects of polarization mode dispersion (PMD). The quantum bit error rate (QBER) was observed to be 5.0\%, with a secure key rate of 589.5 cps achieved through a specially optimized, unbalanced basis selection probability. Additional testing on a 32.5~km fiber spool yielded a key rate of 50.4~bps secure key rate, suggesting the potential for this SPS in polarization QKD over a much longer distance. Our experiment demonstrates the feasibility of implementing GaN-based room-temperature telecom SPSs for polarization QKD in deployed fiber links. 

\section{Results}
\subsection{Single-Photon Generation}
In this experiment, an oil immersion confocal microscope was used to both optically pump (1064 nm) a GaN sample on patterned sapphire substrate (PSS), and collect the luminescence above 1200~nm using appropriate filtering. An overview of the setup layout is depicted in Fig.~\ref{setup}. All the experiments in this work were completed with the sample under ambient laboratory temperature, without any dedicated temperature control. 

\begin{figure}[htbp!]
\centering
\includegraphics[width=0.48\textwidth]
{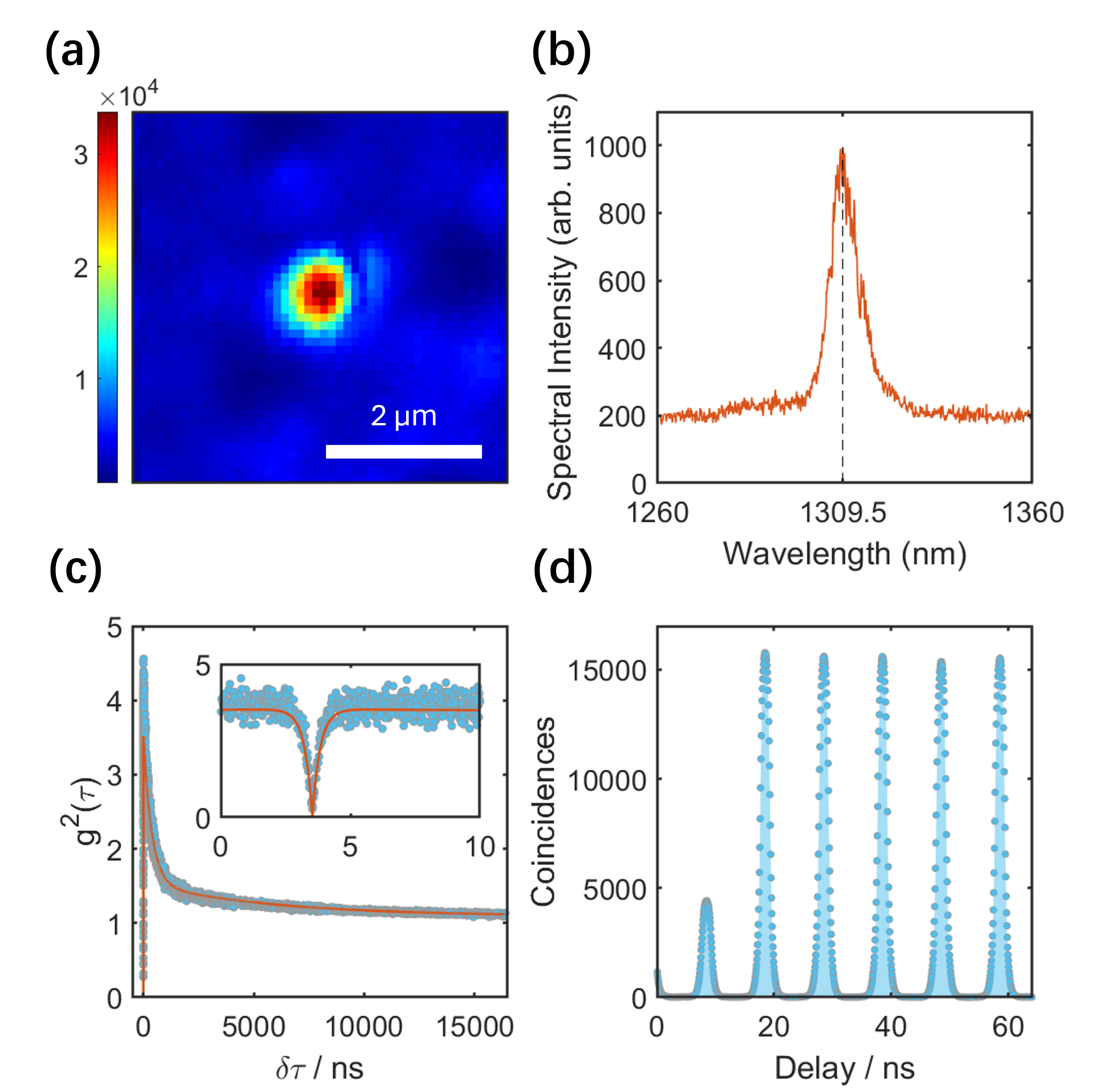}
\centering
\caption{The optical characterization of the GaN SPS emitter sample used in this experiment, taken with a 1200~nm long-pass filter and a 1000 nm dichroic mirror. (a). The spatially resolved confocal map of the emitter. The structure of the patterned sapphire substrate is visible in the background. (b). The photoluminescence spectrum of the sample emitter when the pump wavelength is 1064~nm. (c). The $g^{(2)}(\tau)$ data obtained using 0.1~mW continuous pump (blue dots), fitted with the three-level model (orange line). (d). The measured $g^{(2)}(\tau)$ obtained using a 100~MHz pulsed pump with a width of 80~ps.}
\label{SPS}
\end{figure}

A bright defect-based emitter is located within the GaN layer, with an emission wavelength centered at 1309.5~nm to minimize the loss and dispersion in the telecom fiber. The PSS increases the extraction efficiency of the emitted photons, and therefore the detected count rate~\cite{Yuzhou18}. A confocal map around the emitter and a photolumininescence spectrum are shown in Fig.~\ref{SPS}(a) and (b) respectively.  

To quantify the SPS quality the second-order correlation function $g^{(2)}(\tau)$ was measured under continuous and pulsed conditions. Fig.~\ref{SPS}(c) shows a measured $g^{(2)}(0)$ of $0.28 \pm 0.04$ with continuous pump. When the pump laser was switched to pulsed mode (80~ps duration) the measured pulsed $g^{(2)}(0)$ curve is $0.323~\pm~0.005$ as shown in Fig.~\ref{SPS}~(d). These $g^{2}$ values indicate that the emitter exhibited low multi-photon possibility and that most of the signal will not be lost during the key distillation phase. 

%This source has been used in an experiment that demonstrated time-bin and phase-encoded QKD previously, but the broad emission spectrum (approximately 7~nm) resulted in strong effects from polarization mode dispersion (PMD). In this work, steps were taken to reduce the effects of PMD to demonstrate that the source is compatible with polarization-encoded communication over optical fiber. %Although the principle of the SPS remains unclear thus far, the bunching shoulders near the zero delay area imply more than two energy levels of the emitter\cite{fishman2023photon}. The continuous $g^{(2)}(\tau)$ is fitted with a three-level model: 
%$$g^{(2)}(\tau)=1-\alpha*exp(-|\tau|/\tau_1)+\beta exp(-|\tau|/\tau_2)$$ 
%where $\beta$ relates to the transfer rates between energy levels, $\tau_1$ and $\tau_2$ are the lifetimes of the excited state and the metastable state.  

\subsection{Polarization Mode Dispersion in Quantum Channels}
A 4-state polarization BB84 protocol was implemented in this experiment. Due to the relatively broad bandwidth (7~nm) of our quantum source, polarization mode dispersion (PMD) has been identified as a major factor affecting the QBER. In addition to causing a small difference in the arrival time between different polarization modes, PMD in the quantum channel can have a significant effect on the QBER of polarization protocols - a consequence of changes in the output polarization against wavelength~\cite{gordon2000pmd,yichengthesis,heffner1992automated}. For any small bandwidth where the first-order PMD is dominant, the effect could be visualized as the outgoing state tracing along an arc centered around a rotation axis on the Poincare sphere. A pair of polarization states, known as the Principal States of Polarization (PSPs), can be thus found along the rotation axis. These states remain unaffected by first-order PMD and can be used to minimize its impact in this experiment. The total PMD vector is defined in the format of the Taylor series:
\begin{equation}
\bm\tau(\omega)=\bm{\tau}(\omega_0)+\frac{d\bm{\tau}}{d\omega}\Delta\omega+...
\end{equation}
In the formula, \(\bm{\tau}(\omega_0)\) represents the first-order PMD vector. Its magnitude corresponds to the delay between the two PSPs, or, in other words, the value of the differential group delay, with the direction pointing toward the slower PSP~\cite{Nelson2005}. The higher-order terms in the series describe how the PMD vector varies with wavelength. 

\begin{figure}
\includegraphics[scale=0.4]{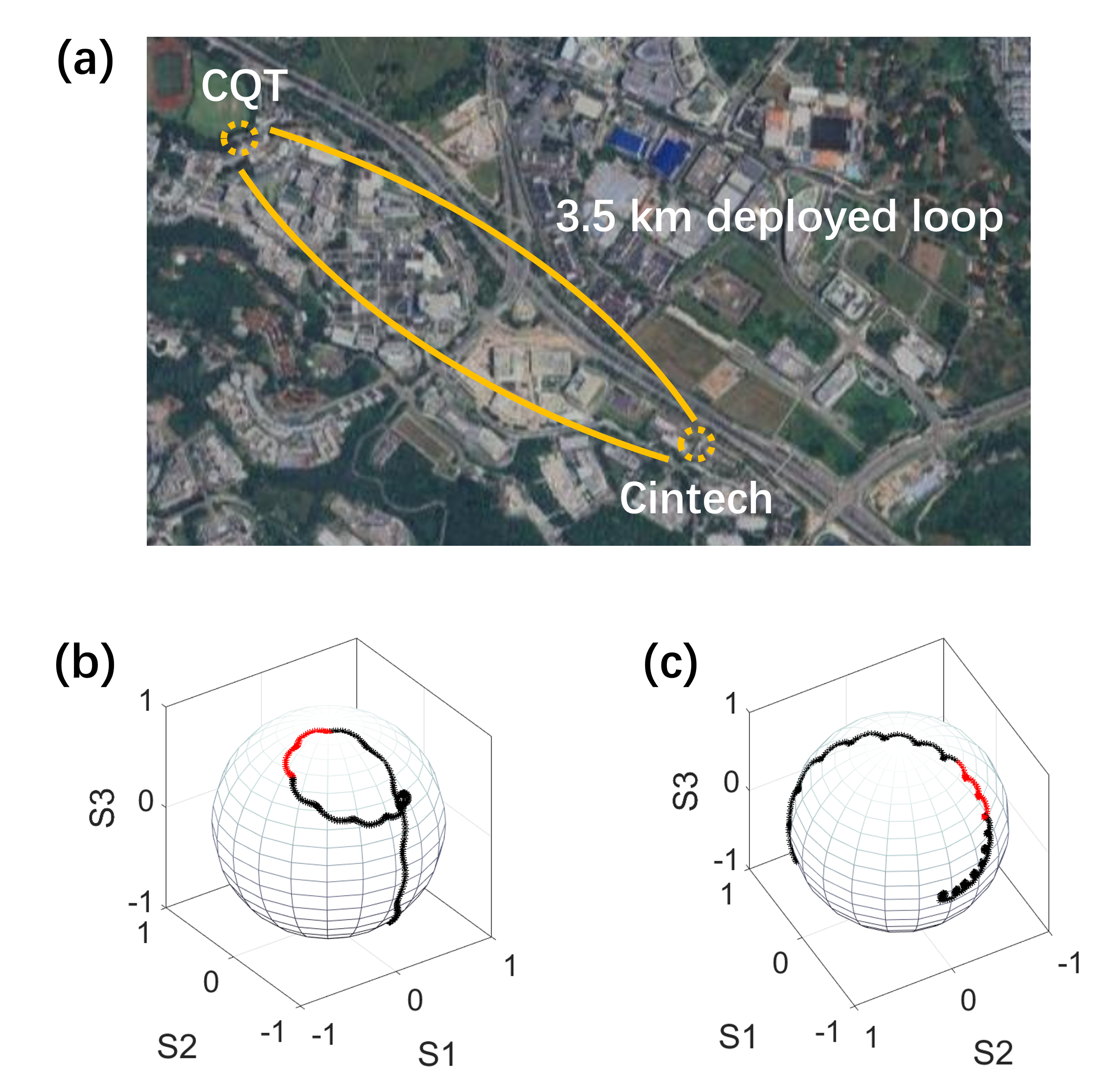}
\centering
\caption{Illustration of the deployed channel and the PMD characterization. (a). Map of the start and end points of the deployed channel. (b) and (c) demonstrate the spread in polarization due to PMD in the deployed channel and spool respectively. The wavelength range was from 1280~nm to 1340~nm. The wavelength band of the photons from the SPS (1309.5~nm $\pm$ 3.5~nm) is marked in red.}
\label{PMD}
\end{figure}

Such distortion from the fiber channel could be visualized by sending a narrow-band tunable laser and measuring the polarization trajectory on the Poincare sphere against the wavelength change. We conducted tests on a 3.5~km deployed fiber as in Fig.~\ref{PMD}~(a) and a 32.5~km fiber spool. The results are plotted in Fig.~\ref{PMD}~(b) and (c), which intuitively depict how the polarization state distributed over long fiber is dispersed in a broad wavelength range (60 nm). The polarization-wavelength trace in deployed fibers typically exhibits more irregular shapes than in spools due to the increased influence of higher-order PMD. The trajectory is also revealed to be reasonably close to an arc within a small wavelength range, where the higher-order terms have a minor effect. 
\begin{figure*}[htbp]
\includegraphics[scale=0.28]{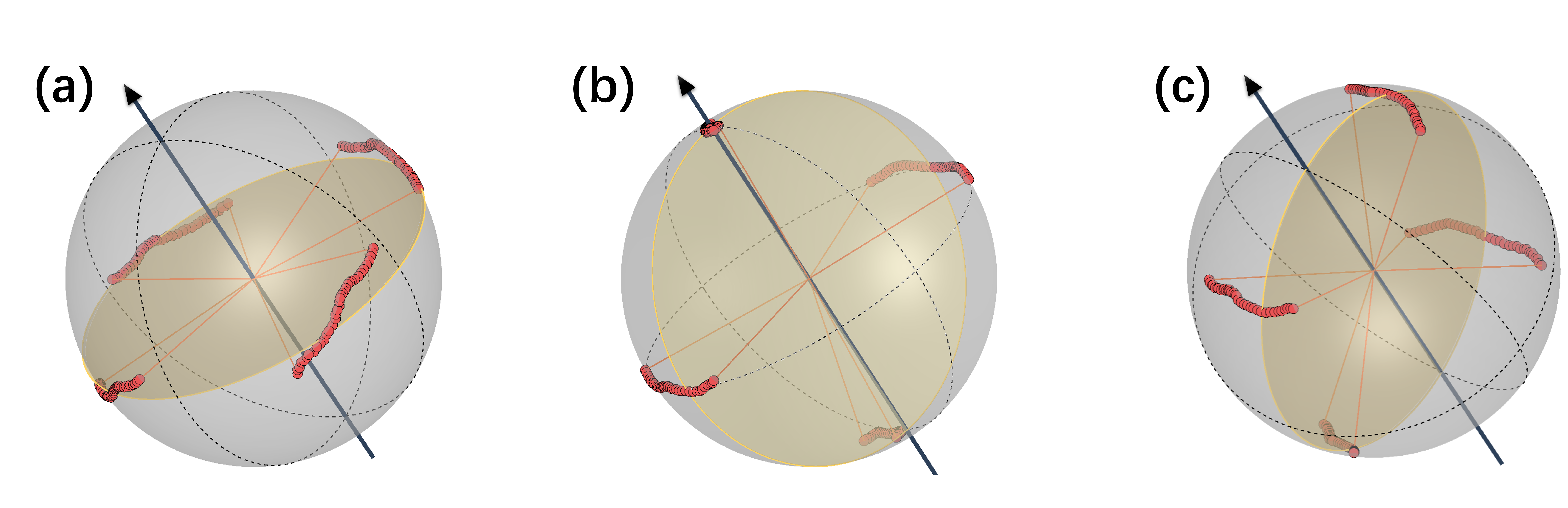}
\centering
\caption{The polarization trajectories (red dots), when starting at different locations on the Poincare sphere relative to the PMD vector (black arrow). This is for a wavelength range of 7~nm and measured after the photons travel over 3.5~km of deployed fiber. Here we show three representative choices for the 4 states used in the BB84 QKD protocol, that can influence the final QBER value. (a). The special case when the 4 states are all orthogonal to the PMD vector. All states suffer a maximum scattering. (b). The case when two of the states are orthogonal to the PMD vector. Only two states experience maximal scattering. (c). All the states have equal but reduced scattering. }
\label{4statePMD}
\end{figure*}
Based on this, it has been demonstrated that by selecting the appropriate basis sets, the impact of PMD can be reduced~\cite{rodimin2024impactpolarizationmodedispersion} by half compared to the worst case. The impact of first-order PMD on the Poincare sphere is described by:
\begin{equation}
\Delta\theta=|\bm\tau|\Delta\omega
\end{equation}
where $\Delta\theta$ is the field angle of the scattering trace. Therefore, a minimum total scattering could be achieved if the PMD vector sits on the same great circle as the four states of polarization QKD. We demonstrated this experimentally in Fig.~\ref{4statePMD}. According to the results, states closer to the PSP experience less impact from PMD. A poor choice, as shown in Fig.~\ref{4statePMD}(a), results in $\Delta\theta$ scatter for all four states, while better choices can reduce the QBER by setting no scatter for two of the states in (b) or $\frac{\Delta\theta}{\sqrt{2}}$ for all of them in (c). Table. \ref{Angle_Arc} indicates the central angles of each arc trajectory in Fig. \ref{4statePMD} for easy comparison. Generally, the 4-state great circle should intersect the PSP vectors. Case (b) is applied in our experiment.  

By observing the scatter trajectory of two sets of orthogonal polarization bases against wavelength as illustrated in Fig.~\ref{4statePMD}, the magnitude of $\Delta\theta$ could be determined. The PMD value of the link could be thus estimated by calculating $\Delta\theta/\Delta\omega$~\cite{williams2004pmd,Jopson1999}. The deployed link exhibited a PMD parameter of $0.46~ps/\sqrt{km}$, which is higher than the $0.13~ps/\sqrt{km}$ of fiber spool and the reference value $0.1$ to $0.2~ps/\sqrt{km}$ for common SMF-28 hence became a major source of our QBER. The difference may be attributed to aging, advancements in manufacturing technology that have reduced PMD, as well as the lower PMD observed in the uncabled fiber compared to the cabled fiber~\cite{ITUG652}.
\begin{table}
    \centering
    \caption{Central angles of each state}
    \label{Angle_Arc}
    \begin{tabular}{cccc}
        \hline
        \textbf{State} & \textbf{Case (a)} & \textbf{Case (b)} & \textbf{Case (c)} \\  \hline
        $|0\rangle$ & 51.5\(^\circ\) & 12.9\(^\circ\) & 39.6\(^\circ\) \\
        $|1\rangle$ & 51.6\(^\circ\) & 14.1\(^\circ\) & 40.8\(^\circ\) \\
        $|+\rangle$ & 50.4\(^\circ\) & 66.5\(^\circ\) & 24.2\(^\circ\) \\
        $|-\rangle$ & 52.8\(^\circ\) & 55.1\(^\circ\) & 59.0\(^\circ\) \\
        \hline
    \end{tabular}
\end{table}

\subsection{Key Rates}
\begin{figure*}[htbp]
\includegraphics[scale=0.33]{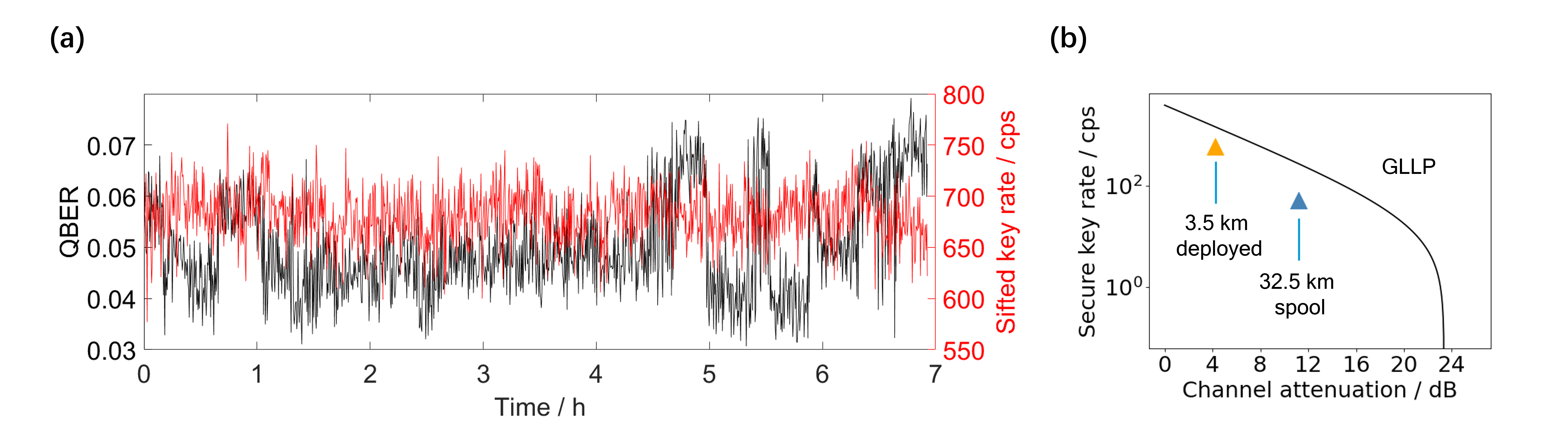} 
\centering
\caption{Results of the QKD experiment over quantum channels. (a). The record of average QBER and sifted key rate over the deployed channel within 7~hours. (b). The experimental results over different quantum channels compared with the GLLP bound based on our current devices. }
\label{QBER_combine}
\end{figure*}

Fig.~\ref{QBER_combine}~(a) records the variation in QBER and sifted key rate for signals transmitted over the deployed fiber, with an integration time of 20~s. The D and A polarizations were pre-aligned to the PSPs of the channel as we established earlier, which resulted in a DA QBER of 1.7\% and LR QBER of 8.3\%. A raw key rate of 1349.6~bps with an average QBER of 5.0\% was obtained over 7 hours without active feedback. With the experimental parameters of the devices and the assumed variables summarized in Table.~\ref{parameters}, a finite-key analysis could be carried out to bound the amount of secure key to be $\epsilon_{sec}$-secret and $\epsilon_{cor}$-correct~\cite{Marco2012tight}. In this experiment, the DA and LR bases had the same preparation and measurement probability, where a secure key rate of 247.3~bps was observed with finite key analysis. However, considering the nature of unbalanced QBER caused by PMD, an unbalanced ratio of bases could be designed to maximize the key rate. A secure key rate of 585.9~bps could be achieved under $p_z=0.997$ and $p_x=0.003$, which well approaches the asymptotic limitation from GLLP~\cite{gottesman2004security, lutkenhaus2000security} considering our channel and device characteristics as shown in Fig.~\ref{QBER_combine}(b). The result for the 32.5 km spool is also indicated with a reasonably low QBER of 3.2\% and a secure key rate of 50.4 bps. 
\begin{table}[!ht]
    \centering
    \caption{Overview of the Parameters}
    \label{parameters}
    \begin{tabular}{lll}
    \hline
        \textbf{Experiment} & \textbf{Parameter} & \textbf{Value} \\ \hline
        System repetition rate & $\nu_{rep}$ & 80 MHz \\ 
        Source efficiency & $r_c$ & $4.19\times 10^{-4}$ \\ 
        Detector overall efficiency & $\eta_{Det}$ & 37.5 \% \\ 
        Dark count rate & $p_{Dark}$ & $1\times 10^{-7}$ \\ 
        QBER @ 0 km & $e_{0}$ &0.9 \% \\
        Channel loss (deployed) & $\l_C$ & 4.0 dB \\
        Channel loss (spool) & $\l_C$ & 11.2 dB \\
        Alice device loss & $l_A$ & 6.2 dB\\
        Bob device loss & $l_B$& 1.7 dB\\ \hline
        \textbf{Simulation} & \textbf{Parameter} & \textbf{Value} \\ \hline
        Security parameter & $\epsilon_{sec}$ & $10^{-12}$\\
        Correction parameter & $\epsilon_{cor}$ & $10^{-12}$\\
        Error correction factor & $f$ & 1.16\\
        \hline
        %Assumed fiber loss & $\alpha$ & 0.4 dB/km \\
        %Assumed security parameter &$\epsilon$&10^{-12}\\
    \end{tabular}
\end{table}

%A QBER of 4.4 \% with a sifted key rate of 411 bps is observed during a 15-minute run. Fig. ??? shows a record of QBER and sifted key rate over ??? hours. With the experimental parameters summarized in Table. \ref{parameters}, a finite-key analysis for the secure key gain in different running time is performed as Fig. ???, which highlights that the stability of our system can overcome the finite size effect and approach the asymptotic limit\cite{gottesman2004security, lutkenhaus2000security}. The experiment over the 32.5 km spool, which generates a 68.9 bps sifted key rate with 3.2 \% QBER, is also posited in the picture. Considering the post-processing and finite size effect\cite{Cai_2009,bunandar2020numerical}, the simulated secure key rate is calculated by:
%$$N_{sec}=(N_{sif}-N_{m})(1-H(e_1))-leak_{EC}-leak_{EV}-\Delta$$
%where $N_{sec} $ is the length of the secure bit string, $N_{sif}$ is the length of the sifted key, $e_1=\frac{e N_{sif}}{N_{sif}-N_{m}}$ is the upper bound of single-photon QBER, $\Delta$ is the extra information loss due to the finite size effect. $leak_{EC}$ represents the information leakage in error correction. $leak_{EV}$ is the information leakage during error verification. $N_{m}$ is the amount of multi-photon events Alice sends into the quantum channel, which is upper bounded by $p_m\leq \frac{g^{(2)}\mu^2}{2}$. The mean photon number $\mu$ is estimated with the PL brightness, repetition rate, and SNSPD efficiency. 

\section{Discussion}
A polarization-encoded BB84 experiment was performed with the single photons generated by the GaN defect SPS, which emits telecom photons under room temperature. With a base-choosing strategy to minimize the PMD impact, a field trial over a 3.5~km deployed fiber with high PMD was demonstrated, where a secure key rate of 585.9~bps with high stability for 7~hours could be achieved according to the finite-key analysis, while the feasibility of longer distribution was confirmed by another experiment over a 32.5~km spool. Both results have approached the GLLP limitation with current devices. The main source of QBER is the depolarization effect caused by PMD due to a relatively large emission spectrum of the SPS.  The fiber spool exhibited substantially lower PMD than the deployed fiber, suggesting that the SPS could also support polarization-encoding over a reasonable fiber distance with reduced PMD. This could be achieved either by narrowing the linewidth of SPS emission~\cite{meunier2023telecom}, switching to fiber models with lower PMD~\cite{rodimin2024impactpolarizationmodedispersion}, or seeking proper PMD-compensating methods ~\cite{Cristianthesis}. The results reveal the potential for room-temperature telecom SPSs to be applied in polarization-encoded QKD, and therefore pave the way for more practical implementations of SPSs on QKD neatly and easily. 
% \usepackage{booktabs}

\section{Materials and Methods}
\subsection{Polarization-encoded BB84 setup}
For state preparation, a Sagnac-like polarization modulator~\cite{li2019high} capable of a high repetition rate was employed, as illustrated in Fig. \ref{setup}. As this is a proof-of-concept experiment, the modulator settings were chosen from a prepared file and not a quantum random number generator. The experiment was synchronized to a clock with an 80~MHz repetition rate. 

The incoming photon wavepacket was separated into $|H\rangle$ and $|V\rangle$ arms by a Polarization Beam Splitter (PBS) with equal probability. These arms were made from polarization-maintaining fiber and arranged such that the photons propagate simultaneously along the fast axis. Since the phase modulator sitted asymmetrically in the loop, the pulsed modulation signal only modified the clockwise or the anti-clockwise signal phase, generating a phase difference between $|H\rangle$ and $|V\rangle$.  After being combined by the PBS, a series of polarization states $|H\rangle+e^{i\phi_V}|V\rangle$ could be created by controlling the electro-optical phase modification $\phi_V$. With $\phi_V$ calibrated to $0, \pi/2, \pi, 3\pi/2$, the polarization states $|D\rangle, |L\rangle, |A\rangle$ and $|R\rangle$ required by the BB84 protocol were generated. 

The receiver was a passive basis-selection polarization measurement setup. A non-polarizing beamsplitter was applied to randomly select a basis for measurement. The single photon signals were measured with superconducting nanowire detectors, and their time of arrival recorded by a time tagger.  
%\subsection{Modulation Control}
%An iXBlue MPX1300-LN-0.1 phase modulator is controlled by an 8-bit digital-analog converter with a range of 0 to 10 V, driven by a Xilinx ZYNQ FPGA chip. A 128 Mb binary file is preloaded into memory to define the sequence for state preparation. A crystal oscillator provides three synchronization signals for the DAC, the pump laser, and the time tagger, respectively.
\subsection{Key Analysis}
The length of the secure key is given by:
\begin{equation}
l=n_zA_z(1-h(Q_x+\delta))-leak_{EC}-log\frac{2}{\epsilon_{sec}^2\epsilon_{cor}}
\end{equation}
The above formula separates the sifted key into bit basis Z with length $n_z$ and phase basis X with length $n_x$ for security analysis, where the Z basis is used to generate the secure key and the X basis is employed to bound Eve's information. Alice and Bob are assumed to agree on choosing the bases with possibilities $p_x$ and $p_z$. $l$ is the length of the secure key, $A_z=1-\frac{p_m}{p_{det}p_z}$ and $A_x=1-\frac{p_m}{p_{det}p_x}$ correspond to the security leakage from the multi-photon emissions in Z basis and X basis respectively, where $p_m$ is the amount of multi-photon events Alice sends into the quantum channel and upper bounded by $p_m\leq \frac{g^{(2)}(0)\mu^2}{2}$, while $p_{det}$ is the detection possibility. The mean photon number $\mu$ is estimated with the PL brightness, repetition rate, and SNSPD efficiency. $h(q)$ represents the binary Shannon entropy function. $Q_x=\frac{e_x}{A_x}$ is the single photon QBER of the X basis assuming all multi-photon states are untrustable, $\delta=\sqrt{\frac{(n_z+n_x)(n_x+1)}{n_zn_x^2}\ln\frac{2}{\epsilon_{sec}}}$ is the statistical fluctuation of QBER estimation, and $\epsilon_{cor}$ is the security parameter in error correction. $e_z$ and $e_x$ are the experimentally observed QBER in Z and X bases. $leak_{EC}=fh(e_z)n_z$ represents the information leakage in error correction. 

\begin{acknowledgments}
We gratefully acknowledge Prof. Scarani Valerio for suggesting and reviewing the finite key analysis method. This work is supported by the National Research Foundation, Singapore and A*STAR under its Quantum Engineering Programme (NRF2021-QEP2-01-P02, National Quantum-Safe Network, NRF2021-QEP2-04-P01, NRF2022-QEP2-02-P13), ASTAR (M21K2c0116, M24M8b0004), Singapore National Research Foundation(NRF-CRP22-2019-0004, NRF2023-ITC004-001, NRFCRP30-2023-0003, NRF-MSG-2023-0002), Singapore Ministry of Education Tier 2 Grant (MOE-T2EP50222-0018). We acknowledge Netlink Trust for the provisioning of the fibre network. 
\end{acknowledgments}

\section*{Author Contributions}
X.Z. and H.Z. performed the experiment and coding and analyzed the data. J. E. and M. M. implemented the methods on the extraction and measurement of the SPS. J. A. G. and C. R. M. developed ideas for the PMD section. W. G. and A. L. supervised all work, as well as contributed to deciding the experiment scheme and setting up the physical models. All authors discussed the results and contributed to the manuscript.
Another experiment was conducted parallelly using a complementary method, verifying the feasibility of time-bin encoding protocols with the GaN SPS.

%\appendix

%\section{Appendixes}
%\section{A little more on appendixes}

\bibliography{apssamp}% Produces the bibliography via BibTeX.

\providecommand{\noopsort}[1]{}\providecommand{\singleletter}[1]{#1}%
\begin{thebibliography}{42}%
\makeatletter
\providecommand \@ifxundefined [1]{%
 \@ifx{#1\undefined}
}%
\providecommand \@ifnum [1]{%
 \ifnum #1\expandafter \@firstoftwo
 \else \expandafter \@secondoftwo
 \fi
}%
\providecommand \@ifx [1]{%
 \ifx #1\expandafter \@firstoftwo
 \else \expandafter \@secondoftwo
 \fi
}%
\providecommand \natexlab [1]{#1}%
\providecommand \enquote  [1]{``#1''}%
\providecommand \bibnamefont  [1]{#1}%
\providecommand \bibfnamefont [1]{#1}%
\providecommand \citenamefont [1]{#1}%
\providecommand \href@noop [0]{\@secondoftwo}%
\providecommand \href [0]{\begingroup \@sanitize@url \@href}%
\providecommand \@href[1]{\@@startlink{#1}\@@href}%
\providecommand \@@href[1]{\endgroup#1\@@endlink}%
\providecommand \@sanitize@url [0]{\catcode `\\12\catcode `\$12\catcode `\&12\catcode `\#12\catcode `\^12\catcode `\_12\catcode `\%12\relax}%
\providecommand \@@startlink[1]{}%
\providecommand \@@endlink[0]{}%
\providecommand \url  [0]{\begingroup\@sanitize@url \@url }%
\providecommand \@url [1]{\endgroup\@href {#1}{\urlprefix }}%
\providecommand \urlprefix  [0]{URL }%
\providecommand \Eprint [0]{\href }%
\providecommand \doibase [0]{https://doi.org/}%
\providecommand \selectlanguage [0]{\@gobble}%
\providecommand \bibinfo  [0]{\@secondoftwo}%
\providecommand \bibfield  [0]{\@secondoftwo}%
\providecommand \translation [1]{[#1]}%
\providecommand \BibitemOpen [0]{}%
\providecommand \bibitemStop [0]{}%
\providecommand \bibitemNoStop [0]{.\EOS\space}%
\providecommand \EOS [0]{\spacefactor3000\relax}%
\providecommand \BibitemShut  [1]{\csname bibitem#1\endcsname}%
\let\auto@bib@innerbib\@empty
%</preamble>
\bibitem [{\citenamefont {Bennett}\ and\ \citenamefont {Brassard}(2014)}]{Bennett2014}%
  \BibitemOpen
  \bibfield  {author} {\bibinfo {author} {\bibfnamefont {C.~H.}\ \bibnamefont {Bennett}}\ and\ \bibinfo {author} {\bibfnamefont {G.}~\bibnamefont {Brassard}},\ }\bibfield  {title} {\bibinfo {title} {Quantum cryptography: Public key distribution and coin tossing},\ }\href {https://doi.org/10.1016/j.tcs.2014.05.025} {\bibfield  {journal} {\bibinfo  {journal} {Theoretical Computer Science}\ }\textbf {\bibinfo {volume} {560}},\ \bibinfo {pages} {7–11} (\bibinfo {year} {2014})}\BibitemShut {NoStop}%
\bibitem [{\citenamefont {Inoue}\ \emph {et~al.}(2003)\citenamefont {Inoue}, \citenamefont {Waks},\ and\ \citenamefont {Yamamoto}}]{inoue2003differential}%
  \BibitemOpen
  \bibfield  {author} {\bibinfo {author} {\bibfnamefont {K.}~\bibnamefont {Inoue}}, \bibinfo {author} {\bibfnamefont {E.}~\bibnamefont {Waks}},\ and\ \bibinfo {author} {\bibfnamefont {Y.}~\bibnamefont {Yamamoto}},\ }\bibfield  {title} {\bibinfo {title} {Differential-phase-shift quantum key distribution using coherent light},\ }\href@noop {} {\bibfield  {journal} {\bibinfo  {journal} {Physical Review A}\ }\textbf {\bibinfo {volume} {68}},\ \bibinfo {pages} {022317} (\bibinfo {year} {2003})}\BibitemShut {NoStop}%
\bibitem [{\citenamefont {Deng}\ and\ \citenamefont {Long}(2004)}]{deng2004bidirectional}%
  \BibitemOpen
  \bibfield  {author} {\bibinfo {author} {\bibfnamefont {F.-G.}\ \bibnamefont {Deng}}\ and\ \bibinfo {author} {\bibfnamefont {G.~L.}\ \bibnamefont {Long}},\ }\bibfield  {title} {\bibinfo {title} {Bidirectional quantum key distribution protocol with practical faint laser pulses},\ }\href@noop {} {\bibfield  {journal} {\bibinfo  {journal} {Physical Review A—Atomic, Molecular, and Optical Physics}\ }\textbf {\bibinfo {volume} {70}},\ \bibinfo {pages} {012311} (\bibinfo {year} {2004})}\BibitemShut {NoStop}%
\bibitem [{\citenamefont {Stucki}\ \emph {et~al.}(2005)\citenamefont {Stucki}, \citenamefont {Brunner}, \citenamefont {Gisin}, \citenamefont {Scarani},\ and\ \citenamefont {Zbinden}}]{stucki2005fast}%
  \BibitemOpen
  \bibfield  {author} {\bibinfo {author} {\bibfnamefont {D.}~\bibnamefont {Stucki}}, \bibinfo {author} {\bibfnamefont {N.}~\bibnamefont {Brunner}}, \bibinfo {author} {\bibfnamefont {N.}~\bibnamefont {Gisin}}, \bibinfo {author} {\bibfnamefont {V.}~\bibnamefont {Scarani}},\ and\ \bibinfo {author} {\bibfnamefont {H.}~\bibnamefont {Zbinden}},\ }\bibfield  {title} {\bibinfo {title} {Fast and simple one-way quantum key distribution},\ }\href@noop {} {\bibfield  {journal} {\bibinfo  {journal} {Applied Physics Letters}\ }\textbf {\bibinfo {volume} {87}} (\bibinfo {year} {2005})}\BibitemShut {NoStop}%
\bibitem [{\citenamefont {Lo}\ \emph {et~al.}(2005)\citenamefont {Lo}, \citenamefont {Ma},\ and\ \citenamefont {Chen}}]{Lo05prl}%
  \BibitemOpen
  \bibfield  {author} {\bibinfo {author} {\bibfnamefont {H.-K.}\ \bibnamefont {Lo}}, \bibinfo {author} {\bibfnamefont {X.}~\bibnamefont {Ma}},\ and\ \bibinfo {author} {\bibfnamefont {K.}~\bibnamefont {Chen}},\ }\bibfield  {title} {\bibinfo {title} {Decoy state quantum key distribution},\ }\href {https://doi.org/10.1103/PhysRevLett.94.230504} {\bibfield  {journal} {\bibinfo  {journal} {Phys. Rev. Lett.}\ }\textbf {\bibinfo {volume} {94}},\ \bibinfo {pages} {230504} (\bibinfo {year} {2005})}\BibitemShut {NoStop}%
\bibitem [{\citenamefont {Wang}(2005)}]{wang2005beating}%
  \BibitemOpen
  \bibfield  {author} {\bibinfo {author} {\bibfnamefont {X.-B.}\ \bibnamefont {Wang}},\ }\bibfield  {title} {\bibinfo {title} {Beating the photon-number-splitting attack in practical quantum cryptography},\ }\href@noop {} {\bibfield  {journal} {\bibinfo  {journal} {Phys. Rev. Lett.}\ }\textbf {\bibinfo {volume} {94}},\ \bibinfo {pages} {230503} (\bibinfo {year} {2005})}\BibitemShut {NoStop}%
\bibitem [{\citenamefont {Hwang}(2003)}]{hwang2003quantum}%
  \BibitemOpen
  \bibfield  {author} {\bibinfo {author} {\bibfnamefont {W.-Y.}\ \bibnamefont {Hwang}},\ }\bibfield  {title} {\bibinfo {title} {Quantum key distribution with high loss: toward global secure communication},\ }\href@noop {} {\bibfield  {journal} {\bibinfo  {journal} {Phys. Rev. Lett.}\ }\textbf {\bibinfo {volume} {91}},\ \bibinfo {pages} {057901} (\bibinfo {year} {2003})}\BibitemShut {NoStop}%
\bibitem [{\citenamefont {Liu}\ \emph {et~al.}(2023)\citenamefont {Liu}, \citenamefont {Zhang}, \citenamefont {Jiang}, \citenamefont {Chen}, \citenamefont {Ma}, \citenamefont {Zhang}, \citenamefont {Pan}, \citenamefont {Dong}, \citenamefont {Xiong}, \citenamefont {Zhang} \emph {et~al.}}]{liu20231002}%
  \BibitemOpen
  \bibfield  {author} {\bibinfo {author} {\bibfnamefont {Y.}~\bibnamefont {Liu}}, \bibinfo {author} {\bibfnamefont {W.-J.}\ \bibnamefont {Zhang}}, \bibinfo {author} {\bibfnamefont {C.}~\bibnamefont {Jiang}}, \bibinfo {author} {\bibfnamefont {J.-P.}\ \bibnamefont {Chen}}, \bibinfo {author} {\bibfnamefont {D.}~\bibnamefont {Ma}}, \bibinfo {author} {\bibfnamefont {C.}~\bibnamefont {Zhang}}, \bibinfo {author} {\bibfnamefont {W.-X.}\ \bibnamefont {Pan}}, \bibinfo {author} {\bibfnamefont {H.}~\bibnamefont {Dong}}, \bibinfo {author} {\bibfnamefont {J.-M.}\ \bibnamefont {Xiong}}, \bibinfo {author} {\bibfnamefont {C.-J.}\ \bibnamefont {Zhang}}, \emph {et~al.},\ }\bibfield  {title} {\bibinfo {title} {1002 km twin-field quantum key distribution with finite-key analysis},\ }\href@noop {} {\bibfield  {journal} {\bibinfo  {journal} {Quantum Frontiers}\ }\textbf {\bibinfo {volume} {2}},\ \bibinfo {pages} {16} (\bibinfo {year} {2023})}\BibitemShut {NoStop}%
\bibitem [{\citenamefont {Lo}\ and\ \citenamefont {Preskill}(2005)}]{lo2005phase}%
  \BibitemOpen
  \bibfield  {author} {\bibinfo {author} {\bibfnamefont {H.-K.}\ \bibnamefont {Lo}}\ and\ \bibinfo {author} {\bibfnamefont {J.}~\bibnamefont {Preskill}},\ }\bibfield  {title} {\bibinfo {title} {Phase randomization improves the security of quantum key distribution},\ }\href@noop {} {\bibfield  {journal} {\bibinfo  {journal} {arXiv preprint quant-ph/0504209}\ } (\bibinfo {year} {2005})}\BibitemShut {NoStop}%
\bibitem [{\citenamefont {Tang}\ \emph {et~al.}(2013)\citenamefont {Tang}, \citenamefont {Yin}, \citenamefont {Ma}, \citenamefont {Fung}, \citenamefont {Liu}, \citenamefont {Yong}, \citenamefont {Chen}, \citenamefont {Peng}, \citenamefont {Chen},\ and\ \citenamefont {Pan}}]{Tang2013}%
  \BibitemOpen
  \bibfield  {author} {\bibinfo {author} {\bibfnamefont {Y.-L.}\ \bibnamefont {Tang}}, \bibinfo {author} {\bibfnamefont {H.-L.}\ \bibnamefont {Yin}}, \bibinfo {author} {\bibfnamefont {X.}~\bibnamefont {Ma}}, \bibinfo {author} {\bibfnamefont {C.-H.~F.}\ \bibnamefont {Fung}}, \bibinfo {author} {\bibfnamefont {Y.}~\bibnamefont {Liu}}, \bibinfo {author} {\bibfnamefont {H.-L.}\ \bibnamefont {Yong}}, \bibinfo {author} {\bibfnamefont {T.-Y.}\ \bibnamefont {Chen}}, \bibinfo {author} {\bibfnamefont {C.-Z.}\ \bibnamefont {Peng}}, \bibinfo {author} {\bibfnamefont {Z.-B.}\ \bibnamefont {Chen}},\ and\ \bibinfo {author} {\bibfnamefont {J.-W.}\ \bibnamefont {Pan}},\ }\bibfield  {title} {\bibinfo {title} {Source attack of decoy-state quantum key distribution using phase information},\ }\href {https://doi.org/10.1103/PhysRevA.88.022308} {\bibfield  {journal} {\bibinfo  {journal} {Phys. Rev. A}\ }\textbf {\bibinfo {volume} {88}},\ \bibinfo {pages} {022308} (\bibinfo {year} {2013})}\BibitemShut {NoStop}%
\bibitem [{\citenamefont {Senellart}\ \emph {et~al.}(2017)\citenamefont {Senellart}, \citenamefont {Solomon},\ and\ \citenamefont {White}}]{senellart2017high}%
  \BibitemOpen
  \bibfield  {author} {\bibinfo {author} {\bibfnamefont {P.}~\bibnamefont {Senellart}}, \bibinfo {author} {\bibfnamefont {G.}~\bibnamefont {Solomon}},\ and\ \bibinfo {author} {\bibfnamefont {A.}~\bibnamefont {White}},\ }\bibfield  {title} {\bibinfo {title} {High-performance semiconductor quantum-dot single-photon sources},\ }\href@noop {} {\bibfield  {journal} {\bibinfo  {journal} {Nature nanotechnology}\ }\textbf {\bibinfo {volume} {12}},\ \bibinfo {pages} {1026} (\bibinfo {year} {2017})}\BibitemShut {NoStop}%
\bibitem [{\citenamefont {Claudon}\ \emph {et~al.}(2010)\citenamefont {Claudon}, \citenamefont {Bleuse}, \citenamefont {Malik}, \citenamefont {Bazin}, \citenamefont {Jaffrennou}, \citenamefont {Gregersen}, \citenamefont {Sauvan}, \citenamefont {Lalanne},\ and\ \citenamefont {G{\'e}rard}}]{claudon2010highly}%
  \BibitemOpen
  \bibfield  {author} {\bibinfo {author} {\bibfnamefont {J.}~\bibnamefont {Claudon}}, \bibinfo {author} {\bibfnamefont {J.}~\bibnamefont {Bleuse}}, \bibinfo {author} {\bibfnamefont {N.~S.}\ \bibnamefont {Malik}}, \bibinfo {author} {\bibfnamefont {M.}~\bibnamefont {Bazin}}, \bibinfo {author} {\bibfnamefont {P.}~\bibnamefont {Jaffrennou}}, \bibinfo {author} {\bibfnamefont {N.}~\bibnamefont {Gregersen}}, \bibinfo {author} {\bibfnamefont {C.}~\bibnamefont {Sauvan}}, \bibinfo {author} {\bibfnamefont {P.}~\bibnamefont {Lalanne}},\ and\ \bibinfo {author} {\bibfnamefont {J.-M.}\ \bibnamefont {G{\'e}rard}},\ }\bibfield  {title} {\bibinfo {title} {A highly efficient single-photon source based on a quantum dot in a photonic nanowire},\ }\href@noop {} {\bibfield  {journal} {\bibinfo  {journal} {Nature Photonics}\ }\textbf {\bibinfo {volume} {4}},\ \bibinfo {pages} {174} (\bibinfo {year} {2010})}\BibitemShut {NoStop}%
\bibitem [{\citenamefont {Maier}\ \emph {et~al.}(2014)\citenamefont {Maier}, \citenamefont {Gold}, \citenamefont {Forchel}, \citenamefont {Gregersen}, \citenamefont {M{\o}rk}, \citenamefont {H{\"o}fling}, \citenamefont {Schneider},\ and\ \citenamefont {Kamp}}]{maier2014bright}%
  \BibitemOpen
  \bibfield  {author} {\bibinfo {author} {\bibfnamefont {S.}~\bibnamefont {Maier}}, \bibinfo {author} {\bibfnamefont {P.}~\bibnamefont {Gold}}, \bibinfo {author} {\bibfnamefont {A.}~\bibnamefont {Forchel}}, \bibinfo {author} {\bibfnamefont {N.}~\bibnamefont {Gregersen}}, \bibinfo {author} {\bibfnamefont {J.}~\bibnamefont {M{\o}rk}}, \bibinfo {author} {\bibfnamefont {S.}~\bibnamefont {H{\"o}fling}}, \bibinfo {author} {\bibfnamefont {C.}~\bibnamefont {Schneider}},\ and\ \bibinfo {author} {\bibfnamefont {M.}~\bibnamefont {Kamp}},\ }\bibfield  {title} {\bibinfo {title} {Bright single photon source based on self-aligned quantum dot--cavity systems},\ }\href@noop {} {\bibfield  {journal} {\bibinfo  {journal} {Optics express}\ }\textbf {\bibinfo {volume} {22}},\ \bibinfo {pages} {8136} (\bibinfo {year} {2014})}\BibitemShut {NoStop}%
\bibitem [{\citenamefont {Birowosuto}\ \emph {et~al.}(2012)\citenamefont {Birowosuto}, \citenamefont {Sumikura}, \citenamefont {Matsuo}, \citenamefont {Taniyama}, \citenamefont {Van~Veldhoven}, \citenamefont {N{\"o}tzel},\ and\ \citenamefont {Notomi}}]{birowosuto2012fast}%
  \BibitemOpen
  \bibfield  {author} {\bibinfo {author} {\bibfnamefont {M.~D.}\ \bibnamefont {Birowosuto}}, \bibinfo {author} {\bibfnamefont {H.}~\bibnamefont {Sumikura}}, \bibinfo {author} {\bibfnamefont {S.}~\bibnamefont {Matsuo}}, \bibinfo {author} {\bibfnamefont {H.}~\bibnamefont {Taniyama}}, \bibinfo {author} {\bibfnamefont {P.~J.}\ \bibnamefont {Van~Veldhoven}}, \bibinfo {author} {\bibfnamefont {R.}~\bibnamefont {N{\"o}tzel}},\ and\ \bibinfo {author} {\bibfnamefont {M.}~\bibnamefont {Notomi}},\ }\bibfield  {title} {\bibinfo {title} {Fast {Purcell-enhanced} single photon source in 1,550-nm telecom band from a resonant quantum dot-cavity coupling},\ }\href@noop {} {\bibfield  {journal} {\bibinfo  {journal} {Scientific reports}\ }\textbf {\bibinfo {volume} {2}},\ \bibinfo {pages} {321} (\bibinfo {year} {2012})}\BibitemShut {NoStop}%
\bibitem [{\citenamefont {Rodiek}\ \emph {et~al.}(2017)\citenamefont {Rodiek}, \citenamefont {Lopez}, \citenamefont {Hofer}, \citenamefont {Porrovecchio}, \citenamefont {Smid}, \citenamefont {Chu}, \citenamefont {Gotzinger}, \citenamefont {Sandoghdar}, \citenamefont {Lindner}, \citenamefont {Becher} \emph {et~al.}}]{rodiek2017experimental}%
  \BibitemOpen
  \bibfield  {author} {\bibinfo {author} {\bibfnamefont {B.}~\bibnamefont {Rodiek}}, \bibinfo {author} {\bibfnamefont {M.}~\bibnamefont {Lopez}}, \bibinfo {author} {\bibfnamefont {H.}~\bibnamefont {Hofer}}, \bibinfo {author} {\bibfnamefont {G.}~\bibnamefont {Porrovecchio}}, \bibinfo {author} {\bibfnamefont {M.}~\bibnamefont {Smid}}, \bibinfo {author} {\bibfnamefont {X.-L.}\ \bibnamefont {Chu}}, \bibinfo {author} {\bibfnamefont {S.}~\bibnamefont {Gotzinger}}, \bibinfo {author} {\bibfnamefont {V.}~\bibnamefont {Sandoghdar}}, \bibinfo {author} {\bibfnamefont {S.}~\bibnamefont {Lindner}}, \bibinfo {author} {\bibfnamefont {C.}~\bibnamefont {Becher}}, \emph {et~al.},\ }\bibfield  {title} {\bibinfo {title} {Experimental realization of an absolute single-photon source based on a single nitrogen vacancy center in a nanodiamond},\ }\href@noop {} {\bibfield  {journal} {\bibinfo  {journal} {Optica}\ }\textbf {\bibinfo {volume} {4}},\ \bibinfo {pages} {71} (\bibinfo {year} {2017})}\BibitemShut {NoStop}%
\bibitem [{\citenamefont {Schroder}\ \emph {et~al.}(2011)\citenamefont {Schroder}, \citenamefont {Schell}, \citenamefont {Kewes}, \citenamefont {Aichele},\ and\ \citenamefont {Benson}}]{schroder2011fiber}%
  \BibitemOpen
  \bibfield  {author} {\bibinfo {author} {\bibfnamefont {T.}~\bibnamefont {Schroder}}, \bibinfo {author} {\bibfnamefont {A.~W.}\ \bibnamefont {Schell}}, \bibinfo {author} {\bibfnamefont {G.}~\bibnamefont {Kewes}}, \bibinfo {author} {\bibfnamefont {T.}~\bibnamefont {Aichele}},\ and\ \bibinfo {author} {\bibfnamefont {O.}~\bibnamefont {Benson}},\ }\bibfield  {title} {\bibinfo {title} {Fiber-integrated diamond-based single photon source},\ }\href@noop {} {\bibfield  {journal} {\bibinfo  {journal} {Nano letters}\ }\textbf {\bibinfo {volume} {11}},\ \bibinfo {pages} {198} (\bibinfo {year} {2011})}\BibitemShut {NoStop}%
\bibitem [{\citenamefont {Senichev}\ \emph {et~al.}(2021)\citenamefont {Senichev}, \citenamefont {Martin}, \citenamefont {Peana}, \citenamefont {Sychev}, \citenamefont {Xu}, \citenamefont {Lagutchev}, \citenamefont {Boltasseva},\ and\ \citenamefont {Shalaev}}]{Seni21}%
  \BibitemOpen
  \bibfield  {author} {\bibinfo {author} {\bibfnamefont {A.}~\bibnamefont {Senichev}}, \bibinfo {author} {\bibfnamefont {Z.~O.}\ \bibnamefont {Martin}}, \bibinfo {author} {\bibfnamefont {S.}~\bibnamefont {Peana}}, \bibinfo {author} {\bibfnamefont {D.}~\bibnamefont {Sychev}}, \bibinfo {author} {\bibfnamefont {X.}~\bibnamefont {Xu}}, \bibinfo {author} {\bibfnamefont {A.~S.}\ \bibnamefont {Lagutchev}}, \bibinfo {author} {\bibfnamefont {A.}~\bibnamefont {Boltasseva}},\ and\ \bibinfo {author} {\bibfnamefont {V.~M.}\ \bibnamefont {Shalaev}},\ }\bibfield  {title} {\bibinfo {title} {Room-temperature single-photon emitters in silicon nitride},\ }\href {https://doi.org/10.1126/sciadv.abj0627} {\bibfield  {journal} {\bibinfo  {journal} {Science Advances}\ }\textbf {\bibinfo {volume} {7}},\ \bibinfo {pages} {eabj0627} (\bibinfo {year} {2021})},\ \Eprint {https://arxiv.org/abs/https://www.science.org/doi/pdf/10.1126/sciadv.abj0627} {https://www.science.org/doi/pdf/10.1126/sciadv.abj0627} \BibitemShut {NoStop}%
\bibitem [{\citenamefont {Murtaza}\ \emph {et~al.}(2023)\citenamefont {Murtaza}, \citenamefont {Colautti}, \citenamefont {Hilke}, \citenamefont {Lombardi}, \citenamefont {Cataliotti}, \citenamefont {Zavatta}, \citenamefont {Bacco},\ and\ \citenamefont {Toninelli}}]{Murtaza:23}%
  \BibitemOpen
  \bibfield  {author} {\bibinfo {author} {\bibfnamefont {G.}~\bibnamefont {Murtaza}}, \bibinfo {author} {\bibfnamefont {M.}~\bibnamefont {Colautti}}, \bibinfo {author} {\bibfnamefont {M.}~\bibnamefont {Hilke}}, \bibinfo {author} {\bibfnamefont {P.}~\bibnamefont {Lombardi}}, \bibinfo {author} {\bibfnamefont {F.~S.}\ \bibnamefont {Cataliotti}}, \bibinfo {author} {\bibfnamefont {A.}~\bibnamefont {Zavatta}}, \bibinfo {author} {\bibfnamefont {D.}~\bibnamefont {Bacco}},\ and\ \bibinfo {author} {\bibfnamefont {C.}~\bibnamefont {Toninelli}},\ }\bibfield  {title} {\bibinfo {title} {Efficient room-temperature molecular single-photon sources for quantum key distribution},\ }\href@noop {} {\bibfield  {journal} {\bibinfo  {journal} {Opt. Express}\ }\textbf {\bibinfo {volume} {31}},\ \bibinfo {pages} {9437} (\bibinfo {year} {2023})}\BibitemShut {NoStop}%
\bibitem [{\citenamefont {Heindel}\ \emph {et~al.}(2012)\citenamefont {Heindel}, \citenamefont {Kessler}, \citenamefont {Rau}, \citenamefont {Schneider}, \citenamefont {F{\"u}rst}, \citenamefont {Hargart}, \citenamefont {Schulz}, \citenamefont {Eichfelder}, \citenamefont {Ro{\ss}bach}, \citenamefont {Nauerth} \emph {et~al.}}]{heindel2012quantum}%
  \BibitemOpen
  \bibfield  {author} {\bibinfo {author} {\bibfnamefont {T.}~\bibnamefont {Heindel}}, \bibinfo {author} {\bibfnamefont {C.~A.}\ \bibnamefont {Kessler}}, \bibinfo {author} {\bibfnamefont {M.}~\bibnamefont {Rau}}, \bibinfo {author} {\bibfnamefont {C.}~\bibnamefont {Schneider}}, \bibinfo {author} {\bibfnamefont {M.}~\bibnamefont {F{\"u}rst}}, \bibinfo {author} {\bibfnamefont {F.}~\bibnamefont {Hargart}}, \bibinfo {author} {\bibfnamefont {W.-M.}\ \bibnamefont {Schulz}}, \bibinfo {author} {\bibfnamefont {M.}~\bibnamefont {Eichfelder}}, \bibinfo {author} {\bibfnamefont {R.}~\bibnamefont {Ro{\ss}bach}}, \bibinfo {author} {\bibfnamefont {S.}~\bibnamefont {Nauerth}}, \emph {et~al.},\ }\bibfield  {title} {\bibinfo {title} {Quantum key distribution using quantum dot single-photon emitting diodes in the red and near infrared spectral range},\ }\href@noop {} {\bibfield  {journal} {\bibinfo  {journal} {New Journal of Physics}\ }\textbf {\bibinfo {volume} {14}},\ \bibinfo {pages} {083001} (\bibinfo {year}
  {2012})}\BibitemShut {NoStop}%
\bibitem [{\citenamefont {Zhang}\ \emph {et~al.}(2024{\natexlab{a}})\citenamefont {Zhang}, \citenamefont {Ding}, \citenamefont {Li}, \citenamefont {Zhang}, \citenamefont {Guo}, \citenamefont {Wang}, \citenamefont {Ning}, \citenamefont {Xu}, \citenamefont {Liu}, \citenamefont {Zhao}, \citenamefont {Zou}, \citenamefont {Wang}, \citenamefont {Cao}, \citenamefont {He}, \citenamefont {Peng}, \citenamefont {Huo}, \citenamefont {Liao}, \citenamefont {Lu}, \citenamefont {Xu},\ and\ \citenamefont {Pan}}]{zhang2024experimental}%
  \BibitemOpen
  \bibfield  {author} {\bibinfo {author} {\bibfnamefont {Y.}~\bibnamefont {Zhang}}, \bibinfo {author} {\bibfnamefont {X.}~\bibnamefont {Ding}}, \bibinfo {author} {\bibfnamefont {Y.}~\bibnamefont {Li}}, \bibinfo {author} {\bibfnamefont {L.}~\bibnamefont {Zhang}}, \bibinfo {author} {\bibfnamefont {Y.-P.}\ \bibnamefont {Guo}}, \bibinfo {author} {\bibfnamefont {G.-Q.}\ \bibnamefont {Wang}}, \bibinfo {author} {\bibfnamefont {Z.}~\bibnamefont {Ning}}, \bibinfo {author} {\bibfnamefont {M.-C.}\ \bibnamefont {Xu}}, \bibinfo {author} {\bibfnamefont {R.-Z.}\ \bibnamefont {Liu}}, \bibinfo {author} {\bibfnamefont {J.-Y.}\ \bibnamefont {Zhao}}, \bibinfo {author} {\bibfnamefont {G.-Y.}\ \bibnamefont {Zou}}, \bibinfo {author} {\bibfnamefont {H.}~\bibnamefont {Wang}}, \bibinfo {author} {\bibfnamefont {Y.}~\bibnamefont {Cao}}, \bibinfo {author} {\bibfnamefont {Y.-M.}\ \bibnamefont {He}}, \bibinfo {author} {\bibfnamefont {C.-Z.}\ \bibnamefont {Peng}}, \bibinfo {author} {\bibfnamefont {Y.-H.}\ \bibnamefont {Huo}}, \bibinfo
  {author} {\bibfnamefont {S.-K.}\ \bibnamefont {Liao}}, \bibinfo {author} {\bibfnamefont {C.-Y.}\ \bibnamefont {Lu}}, \bibinfo {author} {\bibfnamefont {F.}~\bibnamefont {Xu}},\ and\ \bibinfo {author} {\bibfnamefont {J.-W.}\ \bibnamefont {Pan}},\ }\href@noop {} {\bibinfo {title} {Experimental single-photon quantum key distribution surpassing the fundamental coherent-state rate limit}} (\bibinfo {year} {2024}{\natexlab{a}}),\ \Eprint {https://arxiv.org/abs/2406.02045} {arXiv:2406.02045} \BibitemShut {NoStop}%
\bibitem [{\citenamefont {Zahidy}\ \emph {et~al.}(2024)\citenamefont {Zahidy}, \citenamefont {Mikkelsen}, \citenamefont {M{\"u}ller}, \citenamefont {Da~Lio}, \citenamefont {Krehbiel}, \citenamefont {Wang}, \citenamefont {Bart}, \citenamefont {Wieck}, \citenamefont {Ludwig}, \citenamefont {Galili} \emph {et~al.}}]{zahidy2024quantum}%
  \BibitemOpen
  \bibfield  {author} {\bibinfo {author} {\bibfnamefont {M.}~\bibnamefont {Zahidy}}, \bibinfo {author} {\bibfnamefont {M.~T.}\ \bibnamefont {Mikkelsen}}, \bibinfo {author} {\bibfnamefont {R.}~\bibnamefont {M{\"u}ller}}, \bibinfo {author} {\bibfnamefont {B.}~\bibnamefont {Da~Lio}}, \bibinfo {author} {\bibfnamefont {M.}~\bibnamefont {Krehbiel}}, \bibinfo {author} {\bibfnamefont {Y.}~\bibnamefont {Wang}}, \bibinfo {author} {\bibfnamefont {N.}~\bibnamefont {Bart}}, \bibinfo {author} {\bibfnamefont {A.~D.}\ \bibnamefont {Wieck}}, \bibinfo {author} {\bibfnamefont {A.}~\bibnamefont {Ludwig}}, \bibinfo {author} {\bibfnamefont {M.}~\bibnamefont {Galili}}, \emph {et~al.},\ }\bibfield  {title} {\bibinfo {title} {Quantum key distribution using deterministic single-photon sources over a field-installed fibre link},\ }\href@noop {} {\bibfield  {journal} {\bibinfo  {journal} {npj Quantum Information}\ }\textbf {\bibinfo {volume} {10}},\ \bibinfo {pages} {2} (\bibinfo {year} {2024})}\BibitemShut {NoStop}%
\bibitem [{\citenamefont {Morrison}\ \emph {et~al.}(2023)\citenamefont {Morrison}, \citenamefont {Pousa}, \citenamefont {Graffitti}, \citenamefont {Koong}, \citenamefont {Barrow}, \citenamefont {Stoltz}, \citenamefont {Bouwmeester}, \citenamefont {Jeffers}, \citenamefont {Oi}, \citenamefont {Gerardot} \emph {et~al.}}]{morrison2023single}%
  \BibitemOpen
  \bibfield  {author} {\bibinfo {author} {\bibfnamefont {C.~L.}\ \bibnamefont {Morrison}}, \bibinfo {author} {\bibfnamefont {R.~G.}\ \bibnamefont {Pousa}}, \bibinfo {author} {\bibfnamefont {F.}~\bibnamefont {Graffitti}}, \bibinfo {author} {\bibfnamefont {Z.~X.}\ \bibnamefont {Koong}}, \bibinfo {author} {\bibfnamefont {P.}~\bibnamefont {Barrow}}, \bibinfo {author} {\bibfnamefont {N.~G.}\ \bibnamefont {Stoltz}}, \bibinfo {author} {\bibfnamefont {D.}~\bibnamefont {Bouwmeester}}, \bibinfo {author} {\bibfnamefont {J.}~\bibnamefont {Jeffers}}, \bibinfo {author} {\bibfnamefont {D.~K.}\ \bibnamefont {Oi}}, \bibinfo {author} {\bibfnamefont {B.~D.}\ \bibnamefont {Gerardot}}, \emph {et~al.},\ }\bibfield  {title} {\bibinfo {title} {Single-emitter quantum key distribution over 175 km of fibre with optimised finite key rates},\ }\href@noop {} {\bibfield  {journal} {\bibinfo  {journal} {Nature Communications}\ }\textbf {\bibinfo {volume} {14}},\ \bibinfo {pages} {3573} (\bibinfo {year} {2023})}\BibitemShut {NoStop}%
\bibitem [{\citenamefont {Gyger}\ \emph {et~al.}(2022)\citenamefont {Gyger}, \citenamefont {Zeuner}, \citenamefont {Lettner}, \citenamefont {Bensoussan}, \citenamefont {Carln{\"a}s}, \citenamefont {Ekemar}, \citenamefont {Schweickert}, \citenamefont {Hedlund}, \citenamefont {Hammar}, \citenamefont {Nilsson} \emph {et~al.}}]{gyger2022metropolitan}%
  \BibitemOpen
  \bibfield  {author} {\bibinfo {author} {\bibfnamefont {S.}~\bibnamefont {Gyger}}, \bibinfo {author} {\bibfnamefont {K.~D.}\ \bibnamefont {Zeuner}}, \bibinfo {author} {\bibfnamefont {T.}~\bibnamefont {Lettner}}, \bibinfo {author} {\bibfnamefont {S.}~\bibnamefont {Bensoussan}}, \bibinfo {author} {\bibfnamefont {M.}~\bibnamefont {Carln{\"a}s}}, \bibinfo {author} {\bibfnamefont {L.}~\bibnamefont {Ekemar}}, \bibinfo {author} {\bibfnamefont {L.}~\bibnamefont {Schweickert}}, \bibinfo {author} {\bibfnamefont {C.~R.}\ \bibnamefont {Hedlund}}, \bibinfo {author} {\bibfnamefont {M.}~\bibnamefont {Hammar}}, \bibinfo {author} {\bibfnamefont {T.}~\bibnamefont {Nilsson}}, \emph {et~al.},\ }\bibfield  {title} {\bibinfo {title} {Metropolitan single-photon distribution at 1550 nm for random number generation},\ }\href@noop {} {\bibfield  {journal} {\bibinfo  {journal} {Applied Physics Letters}\ }\textbf {\bibinfo {volume} {121}} (\bibinfo {year} {2022})}\BibitemShut {NoStop}%
\bibitem [{\citenamefont {Takemoto}\ \emph {et~al.}(2015)\citenamefont {Takemoto}, \citenamefont {Nambu}, \citenamefont {Miyazawa}, \citenamefont {Sakuma}, \citenamefont {Yamamoto}, \citenamefont {Yorozu},\ and\ \citenamefont {Arakawa}}]{takemoto2015quantum}%
  \BibitemOpen
  \bibfield  {author} {\bibinfo {author} {\bibfnamefont {K.}~\bibnamefont {Takemoto}}, \bibinfo {author} {\bibfnamefont {Y.}~\bibnamefont {Nambu}}, \bibinfo {author} {\bibfnamefont {T.}~\bibnamefont {Miyazawa}}, \bibinfo {author} {\bibfnamefont {Y.}~\bibnamefont {Sakuma}}, \bibinfo {author} {\bibfnamefont {T.}~\bibnamefont {Yamamoto}}, \bibinfo {author} {\bibfnamefont {S.}~\bibnamefont {Yorozu}},\ and\ \bibinfo {author} {\bibfnamefont {Y.}~\bibnamefont {Arakawa}},\ }\bibfield  {title} {\bibinfo {title} {Quantum key distribution over 120 km using ultrahigh purity single-photon source and superconducting single-photon detectors},\ }\href@noop {} {\bibfield  {journal} {\bibinfo  {journal} {Scientific reports}\ }\textbf {\bibinfo {volume} {5}},\ \bibinfo {pages} {14383} (\bibinfo {year} {2015})}\BibitemShut {NoStop}%
\bibitem [{\citenamefont {Ward}\ \emph {et~al.}(2005)\citenamefont {Ward}, \citenamefont {Karimov}, \citenamefont {Unitt}, \citenamefont {Yuan}, \citenamefont {See}, \citenamefont {Gevaux}, \citenamefont {Shields}, \citenamefont {Atkinson},\ and\ \citenamefont {Ritchie}}]{ward2005demand}%
  \BibitemOpen
  \bibfield  {author} {\bibinfo {author} {\bibfnamefont {M.}~\bibnamefont {Ward}}, \bibinfo {author} {\bibfnamefont {O.}~\bibnamefont {Karimov}}, \bibinfo {author} {\bibfnamefont {D.}~\bibnamefont {Unitt}}, \bibinfo {author} {\bibfnamefont {Z.}~\bibnamefont {Yuan}}, \bibinfo {author} {\bibfnamefont {P.}~\bibnamefont {See}}, \bibinfo {author} {\bibfnamefont {D.}~\bibnamefont {Gevaux}}, \bibinfo {author} {\bibfnamefont {A.}~\bibnamefont {Shields}}, \bibinfo {author} {\bibfnamefont {P.}~\bibnamefont {Atkinson}},\ and\ \bibinfo {author} {\bibfnamefont {D.}~\bibnamefont {Ritchie}},\ }\bibfield  {title} {\bibinfo {title} {On-demand single-photon source for 1.3 $\mu$m telecom fiber},\ }\href@noop {} {\bibfield  {journal} {\bibinfo  {journal} {Applied Physics Letters}\ }\textbf {\bibinfo {volume} {86}} (\bibinfo {year} {2005})}\BibitemShut {NoStop}%
\bibitem [{\citenamefont {Zhou}\ \emph {et~al.}(2018)\citenamefont {Zhou}, \citenamefont {Wang}, \citenamefont {Rasmita}, \citenamefont {Kim}, \citenamefont {Berhane}, \citenamefont {Bodrog}, \citenamefont {Adamo}, \citenamefont {Gali}, \citenamefont {Aharonovich},\ and\ \citenamefont {bo~Gao}}]{Yuzhou18}%
  \BibitemOpen
  \bibfield  {author} {\bibinfo {author} {\bibfnamefont {Y.}~\bibnamefont {Zhou}}, \bibinfo {author} {\bibfnamefont {Z.}~\bibnamefont {Wang}}, \bibinfo {author} {\bibfnamefont {A.}~\bibnamefont {Rasmita}}, \bibinfo {author} {\bibfnamefont {S.}~\bibnamefont {Kim}}, \bibinfo {author} {\bibfnamefont {A.}~\bibnamefont {Berhane}}, \bibinfo {author} {\bibfnamefont {Z.}~\bibnamefont {Bodrog}}, \bibinfo {author} {\bibfnamefont {G.}~\bibnamefont {Adamo}}, \bibinfo {author} {\bibfnamefont {A.}~\bibnamefont {Gali}}, \bibinfo {author} {\bibfnamefont {I.}~\bibnamefont {Aharonovich}},\ and\ \bibinfo {author} {\bibfnamefont {W.}~\bibnamefont {bo~Gao}},\ }\bibfield  {title} {\bibinfo {title} {Room temperature solid-state quantum emitters in the telecom range},\ }\href@noop {} {\bibfield  {journal} {\bibinfo  {journal} {Science Advances}\ }\textbf {\bibinfo {volume} {4}},\ \bibinfo {pages} {eaar3580} (\bibinfo {year} {2018})}\BibitemShut {NoStop}%
\bibitem [{\citenamefont {Zhang}\ \emph {et~al.}(2024{\natexlab{b}})\citenamefont {Zhang}, \citenamefont {Zhang}, \citenamefont {Eng}, \citenamefont {Meunier}, \citenamefont {Yang}, \citenamefont {Ling}, \citenamefont {Zuniga-Perez},\ and\ \citenamefont {Gao}}]{haoran2024}%
  \BibitemOpen
  \bibfield  {author} {\bibinfo {author} {\bibfnamefont {H.}~\bibnamefont {Zhang}}, \bibinfo {author} {\bibfnamefont {X.}~\bibnamefont {Zhang}}, \bibinfo {author} {\bibfnamefont {J.}~\bibnamefont {Eng}}, \bibinfo {author} {\bibfnamefont {M.}~\bibnamefont {Meunier}}, \bibinfo {author} {\bibfnamefont {Y.}~\bibnamefont {Yang}}, \bibinfo {author} {\bibfnamefont {A.}~\bibnamefont {Ling}}, \bibinfo {author} {\bibfnamefont {J.}~\bibnamefont {Zuniga-Perez}},\ and\ \bibinfo {author} {\bibfnamefont {W.}~\bibnamefont {Gao}},\ }\href {https://arxiv.org/abs/2409.18502} {\bibinfo {title} {Metropolitan quantum key distribution using a gan-based room-temperature telecommunication single-photon source}} (\bibinfo {year} {2024}{\natexlab{b}}),\ \Eprint {https://arxiv.org/abs/2409.18502} {arXiv:2409.18502 [quant-ph]} \BibitemShut {NoStop}%
\bibitem [{\citenamefont {Zeng}\ \emph {et~al.}(2022)\citenamefont {Zeng}, \citenamefont {Ngyuen}, \citenamefont {Ai}, \citenamefont {Bennet}, \citenamefont {Solntsev}, \citenamefont {Laucht}, \citenamefont {Al-Juboori}, \citenamefont {Toth}, \citenamefont {Mildren}, \citenamefont {Malaney},\ and\ \citenamefont {Aharonovich}}]{Zeng22}%
  \BibitemOpen
  \bibfield  {author} {\bibinfo {author} {\bibfnamefont {H.~Z.~J.}\ \bibnamefont {Zeng}}, \bibinfo {author} {\bibfnamefont {M.~A.~P.}\ \bibnamefont {Ngyuen}}, \bibinfo {author} {\bibfnamefont {X.}~\bibnamefont {Ai}}, \bibinfo {author} {\bibfnamefont {A.}~\bibnamefont {Bennet}}, \bibinfo {author} {\bibfnamefont {A.~S.}\ \bibnamefont {Solntsev}}, \bibinfo {author} {\bibfnamefont {A.}~\bibnamefont {Laucht}}, \bibinfo {author} {\bibfnamefont {A.}~\bibnamefont {Al-Juboori}}, \bibinfo {author} {\bibfnamefont {M.}~\bibnamefont {Toth}}, \bibinfo {author} {\bibfnamefont {R.~P.}\ \bibnamefont {Mildren}}, \bibinfo {author} {\bibfnamefont {R.}~\bibnamefont {Malaney}},\ and\ \bibinfo {author} {\bibfnamefont {I.}~\bibnamefont {Aharonovich}},\ }\bibfield  {title} {\bibinfo {title} {Integrated room temperature single-photon source for quantum key distribution},\ }\href {https://doi.org/10.1364/OL.454450} {\bibfield  {journal} {\bibinfo  {journal} {Opt. Lett.}\ }\textbf {\bibinfo {volume} {47}},\ \bibinfo {pages} {1673}
  (\bibinfo {year} {2022})}\BibitemShut {NoStop}%
\bibitem [{\citenamefont {Gordon}\ and\ \citenamefont {Kogelnik}(2000)}]{gordon2000pmd}%
  \BibitemOpen
  \bibfield  {author} {\bibinfo {author} {\bibfnamefont {J.}~\bibnamefont {Gordon}}\ and\ \bibinfo {author} {\bibfnamefont {H.}~\bibnamefont {Kogelnik}},\ }\bibfield  {title} {\bibinfo {title} {{PMD} fundamentals: Polarization mode dispersion in optical fibers},\ }\href@noop {} {\bibfield  {journal} {\bibinfo  {journal} {Proceedings of the National Academy of Sciences}\ }\textbf {\bibinfo {volume} {97}},\ \bibinfo {pages} {4541} (\bibinfo {year} {2000})}\BibitemShut {NoStop}%
\bibitem [{\citenamefont {Shi}(2022)}]{yichengthesis}%
  \BibitemOpen
  \bibfield  {author} {\bibinfo {author} {\bibfnamefont {Y.}~\bibnamefont {Shi}},\ }\emph {\bibinfo {title} {Polarization-entangled quantum key distribution over telecommunication fibres}},\ \href@noop {} {Ph.D. thesis},\ \bibinfo  {school} {National University of Singapore} (\bibinfo {year} {2022})\BibitemShut {NoStop}%
\bibitem [{\citenamefont {Heffner}(1992)}]{heffner1992automated}%
  \BibitemOpen
  \bibfield  {author} {\bibinfo {author} {\bibfnamefont {B.~L.}\ \bibnamefont {Heffner}},\ }\bibfield  {title} {\bibinfo {title} {Automated measurement of polarization mode dispersion using {Jones} matrix eigenanalysis},\ }\href@noop {} {\bibfield  {journal} {\bibinfo  {journal} {IEEE photonics technology letters}\ }\textbf {\bibinfo {volume} {4}},\ \bibinfo {pages} {1066} (\bibinfo {year} {1992})}\BibitemShut {NoStop}%
\bibitem [{\citenamefont {Nelson}\ and\ \citenamefont {Jopson}(2005)}]{Nelson2005}%
  \BibitemOpen
  \bibfield  {author} {\bibinfo {author} {\bibfnamefont {L.~E.}\ \bibnamefont {Nelson}}\ and\ \bibinfo {author} {\bibfnamefont {R.~M.}\ \bibnamefont {Jopson}},\ }\bibinfo {title} {Introduction to polarization mode dispersion in optical systems},\ in\ \href {https://doi.org/10.1007/0-387-26307-1_1} {\emph {\bibinfo {booktitle} {Polarization Mode Dispersion}}},\ \bibinfo {editor} {edited by\ \bibinfo {editor} {\bibfnamefont {A.}~\bibnamefont {Galtarossa}}\ and\ \bibinfo {editor} {\bibfnamefont {C.~R.}\ \bibnamefont {Menyuk}}}\ (\bibinfo  {publisher} {Springer New York},\ \bibinfo {address} {New York, NY},\ \bibinfo {year} {2005})\ pp.\ \bibinfo {pages} {1--33}\BibitemShut {NoStop}%
\bibitem [{\citenamefont {Rodimin}\ \emph {et~al.}(2024)\citenamefont {Rodimin}, \citenamefont {Kravtsov}, \citenamefont {Chua}, \citenamefont {Santis}, \citenamefont {Ponasenko}, \citenamefont {Kurochkin}, \citenamefont {Ling},\ and\ \citenamefont {Grieve}}]{rodimin2024impactpolarizationmodedispersion}%
  \BibitemOpen
  \bibfield  {author} {\bibinfo {author} {\bibfnamefont {V.}~\bibnamefont {Rodimin}}, \bibinfo {author} {\bibfnamefont {K.}~\bibnamefont {Kravtsov}}, \bibinfo {author} {\bibfnamefont {R.~M.}\ \bibnamefont {Chua}}, \bibinfo {author} {\bibfnamefont {G.~D.}\ \bibnamefont {Santis}}, \bibinfo {author} {\bibfnamefont {A.}~\bibnamefont {Ponasenko}}, \bibinfo {author} {\bibfnamefont {Y.}~\bibnamefont {Kurochkin}}, \bibinfo {author} {\bibfnamefont {A.}~\bibnamefont {Ling}},\ and\ \bibinfo {author} {\bibfnamefont {J.~A.}\ \bibnamefont {Grieve}},\ }\href@noop {} {\bibinfo {title} {Impact of polarization mode dispersion on entangled photon distribution}} (\bibinfo {year} {2024}),\ \Eprint {https://arxiv.org/abs/2408.01754} {arXiv:2408.01754 [quant-ph]} \BibitemShut {NoStop}%
\bibitem [{\citenamefont {Williams}(2004)}]{williams2004pmd}%
  \BibitemOpen
  \bibfield  {author} {\bibinfo {author} {\bibfnamefont {P.}~\bibnamefont {Williams}},\ }\bibfield  {title} {\bibinfo {title} {{PMD} measurement techniques and how to avoid the pitfalls},\ }\href@noop {} {\bibfield  {journal} {\bibinfo  {journal} {Journal of Optical and Fiber Communications Reports}\ }\textbf {\bibinfo {volume} {1}},\ \bibinfo {pages} {84} (\bibinfo {year} {2004})}\BibitemShut {NoStop}%
\bibitem [{\citenamefont {Jopson}\ \emph {et~al.}(1999)\citenamefont {Jopson}, \citenamefont {Nelson},\ and\ \citenamefont {Kogelnik}}]{Jopson1999}%
  \BibitemOpen
  \bibfield  {author} {\bibinfo {author} {\bibfnamefont {R.}~\bibnamefont {Jopson}}, \bibinfo {author} {\bibfnamefont {L.}~\bibnamefont {Nelson}},\ and\ \bibinfo {author} {\bibfnamefont {H.}~\bibnamefont {Kogelnik}},\ }\bibfield  {title} {\bibinfo {title} {Measurement of second-order polarization-mode dispersion vectors in optical fibers},\ }\href {https://doi.org/10.1109/68.784234} {\bibfield  {journal} {\bibinfo  {journal} {IEEE Photonics Technology Letters}\ }\textbf {\bibinfo {volume} {11}},\ \bibinfo {pages} {1153} (\bibinfo {year} {1999})}\BibitemShut {NoStop}%
\bibitem [{\citenamefont {standardization sector~of ITU}(2016)}]{ITUG652}%
  \BibitemOpen
  \bibfield  {author} {\bibinfo {author} {\bibfnamefont {T.}~\bibnamefont {standardization sector~of ITU}},\ }\bibfield  {title} {\bibinfo {title} {{ITU-T G.652}},\ }\href@noop {} {\bibfield  {journal} {\bibinfo  {journal} {International Telecommunication Union}\ } (\bibinfo {year} {2016})}\BibitemShut {NoStop}%
\bibitem [{\citenamefont {Tomamichel}\ \emph {et~al.}(2012)\citenamefont {Tomamichel}, \citenamefont {Lim}, \citenamefont {Gisin},\ and\ \citenamefont {Renner}}]{Marco2012tight}%
  \BibitemOpen
  \bibfield  {author} {\bibinfo {author} {\bibfnamefont {M.}~\bibnamefont {Tomamichel}}, \bibinfo {author} {\bibfnamefont {C.~C.~W.}\ \bibnamefont {Lim}}, \bibinfo {author} {\bibfnamefont {N.}~\bibnamefont {Gisin}},\ and\ \bibinfo {author} {\bibfnamefont {R.}~\bibnamefont {Renner}},\ }\bibfield  {title} {\bibinfo {title} {Tight finite-key analysis for quantum cryptography},\ }\href@noop {} {\bibfield  {journal} {\bibinfo  {journal} {Nature communications}\ }\textbf {\bibinfo {volume} {3}},\ \bibinfo {pages} {634} (\bibinfo {year} {2012})}\BibitemShut {NoStop}%
\bibitem [{\citenamefont {Gottesman}\ \emph {et~al.}(2004)\citenamefont {Gottesman}, \citenamefont {Lo}, \citenamefont {Lutkenhaus},\ and\ \citenamefont {Preskill}}]{gottesman2004security}%
  \BibitemOpen
  \bibfield  {author} {\bibinfo {author} {\bibfnamefont {D.}~\bibnamefont {Gottesman}}, \bibinfo {author} {\bibfnamefont {H.-K.}\ \bibnamefont {Lo}}, \bibinfo {author} {\bibfnamefont {N.}~\bibnamefont {Lutkenhaus}},\ and\ \bibinfo {author} {\bibfnamefont {J.}~\bibnamefont {Preskill}},\ }\bibfield  {title} {\bibinfo {title} {Security of quantum key distribution with imperfect devices},\ }in\ \href@noop {} {\emph {\bibinfo {booktitle} {International Symposium onInformation Theory, 2004. ISIT 2004. Proceedings.}}}\ (\bibinfo {organization} {IEEE},\ \bibinfo {year} {2004})\ p.\ \bibinfo {pages} {136}\BibitemShut {NoStop}%
\bibitem [{\citenamefont {L{\"u}tkenhaus}(2000)}]{lutkenhaus2000security}%
  \BibitemOpen
  \bibfield  {author} {\bibinfo {author} {\bibfnamefont {N.}~\bibnamefont {L{\"u}tkenhaus}},\ }\bibfield  {title} {\bibinfo {title} {Security against individual attacks for realistic quantum key distribution},\ }\href@noop {} {\bibfield  {journal} {\bibinfo  {journal} {Physical Review A}\ }\textbf {\bibinfo {volume} {61}},\ \bibinfo {pages} {052304} (\bibinfo {year} {2000})}\BibitemShut {NoStop}%
\bibitem [{\citenamefont {Meunier}\ \emph {et~al.}(2023)\citenamefont {Meunier}, \citenamefont {Eng}, \citenamefont {Mu}, \citenamefont {Chenot}, \citenamefont {Br{\"a}ndli}, \citenamefont {de~Mierry}, \citenamefont {Gao},\ and\ \citenamefont {Z{\'u}{\~n}iga-P{\'e}rez}}]{meunier2023telecom}%
  \BibitemOpen
  \bibfield  {author} {\bibinfo {author} {\bibfnamefont {M.}~\bibnamefont {Meunier}}, \bibinfo {author} {\bibfnamefont {J.~J.}\ \bibnamefont {Eng}}, \bibinfo {author} {\bibfnamefont {Z.}~\bibnamefont {Mu}}, \bibinfo {author} {\bibfnamefont {S.}~\bibnamefont {Chenot}}, \bibinfo {author} {\bibfnamefont {V.}~\bibnamefont {Br{\"a}ndli}}, \bibinfo {author} {\bibfnamefont {P.}~\bibnamefont {de~Mierry}}, \bibinfo {author} {\bibfnamefont {W.}~\bibnamefont {Gao}},\ and\ \bibinfo {author} {\bibfnamefont {J.}~\bibnamefont {Z{\'u}{\~n}iga-P{\'e}rez}},\ }\bibfield  {title} {\bibinfo {title} {Telecom single-photon emitters in {GaN} operating at room temperature: embedment into bullseye antennas},\ }\href@noop {} {\bibfield  {journal} {\bibinfo  {journal} {Nanophotonics}\ }\textbf {\bibinfo {volume} {12}},\ \bibinfo {pages} {1405} (\bibinfo {year} {2023})}\BibitemShut {NoStop}%
\bibitem [{\citenamefont {Czegledi}(2018)}]{Cristianthesis}%
  \BibitemOpen
  \bibfield  {author} {\bibinfo {author} {\bibfnamefont {C.~B.}\ \bibnamefont {Czegledi}},\ }\emph {\bibinfo {title} {Modeling and Compensation of Polarization Effects in Fiber-Optic Communication Systems}},\ \href@noop {} {Ph.D. thesis},\ \bibinfo  {school} {CHALMERS UNIVERSITY OF TECHNOLOGY} (\bibinfo {year} {2018})\BibitemShut {NoStop}%
\bibitem [{\citenamefont {Li}\ \emph {et~al.}(2019)\citenamefont {Li}, \citenamefont {Li}, \citenamefont {Xie}, \citenamefont {Li}, \citenamefont {Jiang}, \citenamefont {Cai}, \citenamefont {Ren}, \citenamefont {Yin}, \citenamefont {Liao},\ and\ \citenamefont {Peng}}]{li2019high}%
  \BibitemOpen
  \bibfield  {author} {\bibinfo {author} {\bibfnamefont {Y.}~\bibnamefont {Li}}, \bibinfo {author} {\bibfnamefont {Y.-H.}\ \bibnamefont {Li}}, \bibinfo {author} {\bibfnamefont {H.-B.}\ \bibnamefont {Xie}}, \bibinfo {author} {\bibfnamefont {Z.-P.}\ \bibnamefont {Li}}, \bibinfo {author} {\bibfnamefont {X.}~\bibnamefont {Jiang}}, \bibinfo {author} {\bibfnamefont {W.-Q.}\ \bibnamefont {Cai}}, \bibinfo {author} {\bibfnamefont {J.-G.}\ \bibnamefont {Ren}}, \bibinfo {author} {\bibfnamefont {J.}~\bibnamefont {Yin}}, \bibinfo {author} {\bibfnamefont {S.-K.}\ \bibnamefont {Liao}},\ and\ \bibinfo {author} {\bibfnamefont {C.-Z.}\ \bibnamefont {Peng}},\ }\bibfield  {title} {\bibinfo {title} {High-speed robust polarization modulation for quantum key distribution},\ }\href@noop {} {\bibfield  {journal} {\bibinfo  {journal} {Optics Letters}\ }\textbf {\bibinfo {volume} {44}},\ \bibinfo {pages} {5262} (\bibinfo {year} {2019})}\BibitemShut {NoStop}%
\end{thebibliography}%

\end{document}